\documentclass[prX,twocolumn,showpacs,superscriptaddress,aps]{revtex4}

\usepackage{amsfonts,xfrac}
\usepackage{amsmath}
\usepackage{amssymb}
\usepackage{graphicx}
\usepackage{dcolumn}
\usepackage{dsfont}
\usepackage{times}
\usepackage{epsfig,color}
\usepackage[]{units}
\usepackage{soul}

\usepackage[hang,nooneline]{subfigure}                         
\newcounter{fig}

\def\epsilon{\varepsilon}

\begin{document}
\title{Formation of rarefaction waves in origami-based metamaterials}

\author{H. Yasuda}
\affiliation{Aeronautics \& Astronautics, University of Washington, Seattle, WA 98195-2400, USA}
\author{C. Chong}
\affiliation{Department of Mechanical and Process Engineering (D-MAVT),%
Swiss Federal Institute of Technology (ETH), 8092 Z\"urich, Switzerland}
\affiliation{Department of Mathematics, Bowdoin College, Brunswick, ME 04011, USA}
\author{E. G. Charalampidis}
\affiliation{Department of Mathematics and Statistics, University of Massachusetts, Amherst, MA 01003-4515, USA}
\author{P. G. Kevrekidis}
\affiliation{Department of Mathematics and Statistics, University of Massachusetts, Amherst, MA 01003-4515, USA}
\affiliation{Center for Nonlinear Studies and Theoretical Division, Los Alamos
National Laboratory, Los Alamos, NM 87544, USA}
\author{J. Yang\footnote{Email: jkyang@aa.washington.edu}}
\affiliation{Aeronautics \& Astronautics, University of Washington, Seattle, WA 98195-2400, USA}

\date{\today}

\begin{abstract}
{We investigate the nonlinear wave dynamics of origami-based metamaterials
composed of Tachi-Miura polyhedron (TMP) unit cells. These cells exhibit
strain softening behavior under compression, which can be tuned by modifying
their geometrical configurations or initial folded conditions. 
We assemble these TMP cells into a cluster of origami-based metamaterials, and we theoretically model and numerically analyze their wave transmission
mechanism under external impact.  
Numerical simulations show that origami-based metamaterials can provide a prototypical
platform for the formation of 
 nonlinear 
coherent structures 
in the form of rarefaction waves, which feature a  tensile wavefront upon 
the application of compression to the system. We also demonstrate the existence of
numerically exact traveling rarefaction waves. 
Origami-based metamaterials can be highly useful for 
mitigating shock waves, potentially enabling a wide variety of engineering 
applications.}
\end{abstract}

\maketitle

\section{Introduction}
Recently, origami has attracted a significant amount of attention from 
researchers due to its unique mechanical properties. The most evident 
one is its compactness and deployability, which enables various types 
of expandable engineering structures, e.g., space solar sails~\cite{Mori, Tsuda} 
and solar arrays~\cite{Zirbel}.  Biological 
systems also exploit such compact origami patterns, such as foldable 
tree leaves for metabolic purposes~\cite{Kobayashi}
and stent grafts \cite{You}.
 Another useful aspect of origami-based structures is that origami patterns can enhance
static mechanical properties of structures. For instance, structural bending
rigidity for thin-walled cylindrical structures can be significantly improved
by imposing origami-patterns~\cite{Miura1970}. These origami patterns are
used not only for space structures, but also in commercial products (e.g., 
beverage cans) in order to reduce the thickness of 
thin-walled structures without sacrificing their buckling strength~\cite{Kirin}.

Within the considerable
progress made in the mechanics of origami-based structures, however,
the
primary focus has been placed on the static or quasi-static properties of
origami. For example, recent studies attempted to fabricate origami-based
metamaterials with an eye towards investigating the deployable, auxetic, 
and bistable nature of origami structures~\cite{Schenk, Wei, Cheung2014, Yasuda2015}. 
Limited work has been reported on the impact response of origami-based 
structures~\cite{Schenk2014}, and their wave dynamics is relatively unexplored. 
Plausibly, this lack of studies on the dynamics of origami-based structures 
can be attributed to
the intrinsic characteristic of typical origami structures, which exhibit
limited degrees of freedom (DOF) during their folding/unfolding motions. 
This is particularly true for rigid origami, in which the deformation takes
place only along crease lines while origami facets remain rigid in dynamic conditions. The rigid origami features single-DOF motions 
ideally, and thus, the studies on their wave dynamics have been more or less absent 
under this rigid foldability assumption.
\begin{figure}[htbp]
\centerline{ \includegraphics[width=.475\textwidth]{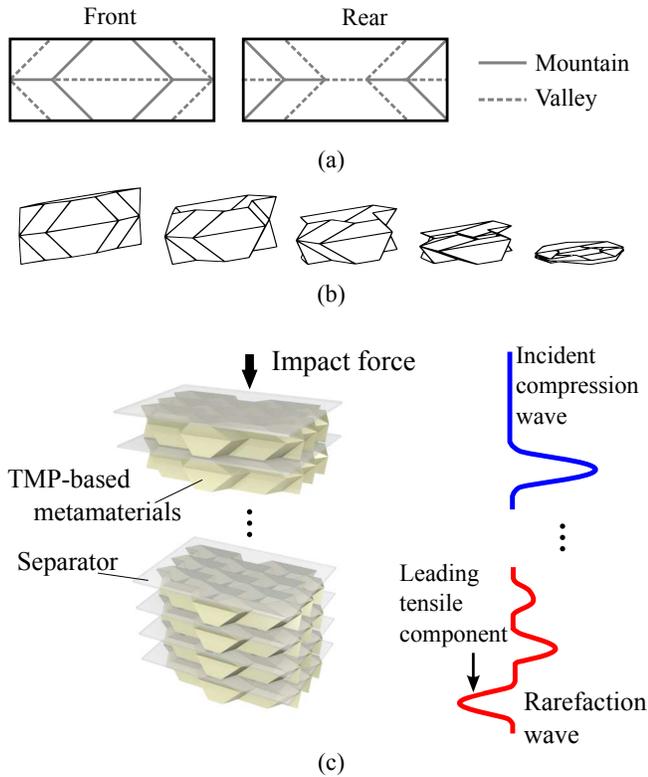}}
\caption{(Color online) \textbf{(a)} Flat front and rear sheets of the TMP 
with mountain (solid lines) and valley (dashed lines) crease lines. \textbf{(b)} 
Folding motion of the TMP unit cell. \textbf{(c)} System consisting of TMP-based 
metamaterials and rigid separators stacked vertically. Each layer 
consists of nine inter-linked TMP unit cells. Conceptual illustrations of incident 
compressive waves and transmitted rarefaction waves are also shown.}
\label{fig:TMP_metamaterials}
\end{figure}

In this study, we use a single-DOF rigid origami structure as a building 
block to assemble multi-DOF mechanical metamaterials, and analyze their 
nonlinear wave dynamics through analytical and numerical approaches. Specifically, 
we employ the Tachi-Miura polyhedron (TMP)~\cite{Miura, Tachi} as a unit cell of the metamaterial
as shown in Fig.~\ref{fig:TMP_metamaterials}. The TMP cell
is made of two adjoined sheets (Fig.~\ref{fig:TMP_metamaterials}(a)), and 
changes its shape from a vertically standing planar body to a horizontally 
flattened one while taking up a finite volume between the two phases (Fig.%
~\ref{fig:TMP_metamaterials}(b)). This volumetric behavior is in contrast to
conventional origami-patterns that feature planar architectures and in-plane
motions (e.g., Miura-ori sheets~ \cite{Miura1985}). In this study, we first 
characterize the kinematics of the TMP cell, showing that it exhibits controllable
strain-softening behavior. 
By cross-linking these TMP unit cells in a horizontal layer and stacking
them up vertically with separators, we form a multi-DOF metamaterial as shown
in Fig.~\ref{fig:TMP_metamaterials}(c). We then conduct analytical and numerical studies to verify that these multi-DOF 
origami structures can support a nonlinear stress wave
in the form of a so-called rarefaction wave, owing to the 
strain softening nature of the assembled 
structure. 

 The rarefaction wave, which can be viewed as an acoustic variant of a depression wave \cite{Falcon},
 has been studied in various settings, including systems of conservation laws \cite{Lax}. More recently, it was proposed in the context of 
discrete systems with strain-softening 
behavior \cite{Herbold,Nesterenko}.
Interestingly, these rarefaction waves feature tensile wavefronts 
despite the application of compressive stresses upon external impact (see the
conceptual illustrations in Fig.~\ref{fig:TMP_metamaterials}(c)). 
In that light, they are fundamentally different from
the commonly encountered dynamical response of nonlinear elastic 
chains which support weakly or even strongly nonlinear
traveling compression waves~\cite{Nesterenko,review_Theo,review_Sen,review_PK}.
Recently, in a quite different
setting of tensegrity structures, rarefaction waves have been identified computationally in the 
elastic softening regime~\cite{fratern}.

Our main scope within the present work is to verify 
the formation and propagation of rarefaction waves in origami-based metamaterials
via two simplified models: a multi-bar linkage model and a lumped mass model. In 
both cases, we confirm that the origami structure disintegrates strong impact
excitations
by forming rarefaction waves, followed by other dispersive wave patterns 
to be discussed in more detail below. We also validate the nonlinear 
nature of the stress waves by calculating the variations of wave speed as a 
function of external force amplitude. Notably, we observe the reduction of 
wave speed as the excitation amplitude increases, which is in sharp contrast
to conventional nonlinear waves seen in nature or engineered 
systems~\cite{Nesterenko}.
In the case of the lumped mass model, we find numerically 
exact traveling waves. We provide a precise characterization of the 
wave speed and amplitude relationship and a way to evaluate the robustness of 
the rarefaction waves through
dynamical stability computations. The findings in this study provide a 
foundation for building a new type of impact 
mitigating structure with tunable characteristics, which does not rely on material
damping or plastic deformation. This study also offers a platform for exciting the rarefaction pulse -- a
far less explored type of traveling wave -- and
examining its characteristics in considerable detail.

The Manuscript is structured as follows: In Sec.~\ref{sec_II}, we describe the 
two simple models of origami-based metamaterials: 
the multi-bar linkage model and the
lumped mass model. In Sec.~\ref{sec_III}, we conduct numerical simulations of
wave propagation upon impact on the chain boundary and compare the wave 
dynamics obtained from these two models. 
Then, in Sec.~\ref{sec_IV} we find numerically exact rarefaction waves
of the lumped mass model. 
Lastly, concluding remarks and future 
work are given in Sec.~\ref{sec_V}.

\begin{figure*}[htbp]
\centerline{ \epsfig{file=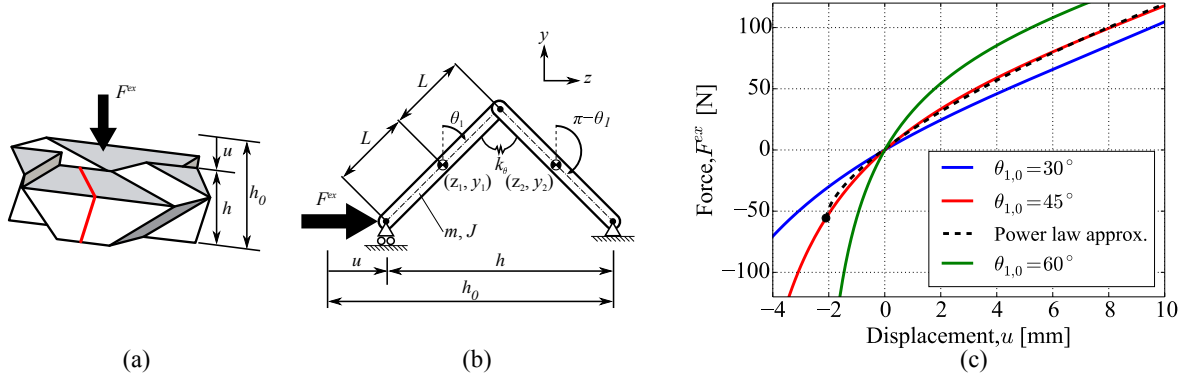,width=0.9 \textwidth} }
\caption{(Color online) \textbf{(a)} TMP
unit cell. \textbf{(b)} Two-bar linkage model representing the folding motion of
the two facets as marked in red lines in \textbf{(a)}. \textbf{(c)} Force-displacement 
relationship of the TMP unit cell with $L=5\,\textrm{mm}$, $k_{\theta}=1.0\,\textrm{Nm/rad}$,
and different initial folding angles: $\theta_{1,0}=30^{\circ}$,  $45^{\circ}$, and
$60^{\circ}$. Dashed line indicates a power law approximation of $\theta_{1,0}=45^{\circ}$ 
case.} 
\label{fig:SingleTMPcell}
\end{figure*}

\section{Modeling of Origami-based Structures}\label{sec_II}
%

%

\subsection{Multi-bar Linkage Model} \label{sec:linkagemodel}
We begin by 
modeling a single TMP cell as shown in Fig.~\ref{fig:SingleTMPcell}. 
For the sake of simplicity, we focus on the folding motion of two adjacent 
facets along the horizontal crease line as marked by the red line in 
Fig.~\ref{fig:SingleTMPcell}(a). Preserving the key features of the TMP, 
such as rigid foldability and single-DOF mobility, we can model the 
folding/unfolding motion of the origami facets into a simple 1D linkage 
model as shown in Fig.~\ref{fig:SingleTMPcell}(b). Here, the unit cell 
consists of two rigid bars (mass $m$ and length $2L$), and the center-of-mass
coordinates of those two bars are ($z_1$, $y_1$, $\theta_1$) and ($z_2$, $y_2$, $\theta_2$). 
The hinge that connects the two bars is equipped with a linear torsional
spring with the torsion coefficient $k_\theta$. The left end of the linkage
structure is supported by a roller joint, which is allowed to move only 
along the \textit{z}-axis up on the application of external force ${F}^{ex}$. 
The right end is fixed by a pin joint. Therefore, the inclined angle of 
the linkage, $\theta_1$, is the only parameter required to describe the 
motion of this unit-cell system. This corresponds to the single-DOF nature
of the TMP cell. 

By using the principle of virtual power \cite{Moon}, we derive the following
equation of motion (see Supplemental Material for details~\cite{SI}): 
\begin{equation} \label{eq:EqMo_single}
\begin{split}
 \left( m{{L}^{2}}/2+J/2+2m{{L}^{2}}{{\cos }^{2}}{{\theta }_{1}} \right){{{\ddot{\theta }}}_{1}}-m{{L}^{2}}\dot{\theta }_{1}^{2}\sin 2{{\theta }_{1}} \\  
 +{{k}_{\theta }}\left( {{\theta }_{1}}-{{\theta }_{1,0}} \right)=-{{F}^{ex}}L\cos {{\theta }_{1}}.
\end{split}
\end{equation}
Here $J$ is the bar's moment of inertia ($J = \frac{mL^{2}}{3}$), and $\theta_{1,0}$
is the initial folding angle (i.e., no torque applied at the hinge in this
initial angle). In the quasi-static case (i.e., acceleration and velocity 
terms are much smaller compared to the external excitation and spring force
terms), we obtain the force-displacement relationship as follows:
\begin{equation} \label{eq:FS_curve}
F^{ex} = -\frac{k_{\theta} \left( \theta_1-\theta_{1,0} \right)}{L\cos{\theta_1}}.
\end{equation}
Using Eq.~\eqref{eq:FS_curve} and the axial displacement expression 
$u = 4L(\sin{\theta_{1,0}} - \sin{\theta_1})$, we can calculate the 
force-displacement response as shown in Fig.~\ref{fig:SingleTMPcell}(c).
We observe that the system exhibits strain
softening behavior in the compressive region, whereas the system shows 
strain hardening response in the tensile domain. Also, it is interesting
to find that this strain softening/hardening behavior can be tuned by controlling
the initial folding angle, $\theta_{1,0}$.

\begin{figure}[htbp]
\centerline{ \epsfig{file=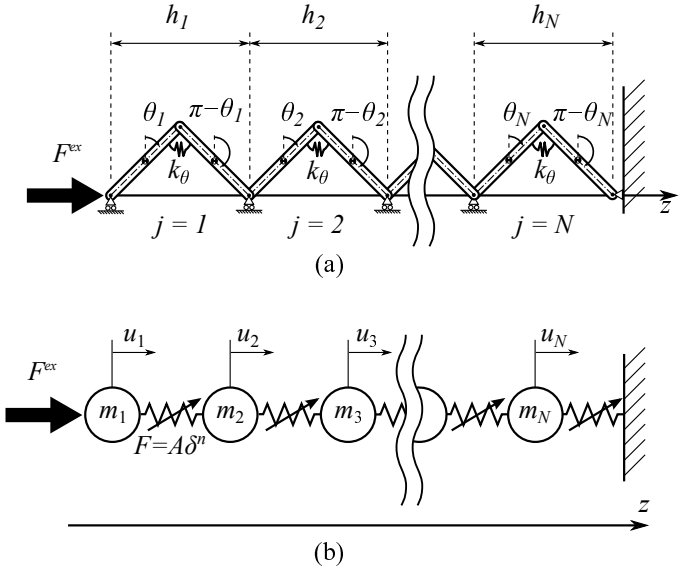,width=0.5 \textwidth} } 
\caption{Schematic illustrations of \textbf{(a)} Multi-bar linkage model and \textbf{(b)} Lumped mass model.}
\label{fig:Modeling_origami}
\end{figure}

 Based on the kinematics of the single unit cell as expressed in Eq.~\eqref{eq:EqMo_single}, 
 we model a chain of $N$-TMP cells as shown in Fig.~\ref{fig:Modeling_origami}(a). 
 In this model, each unit cell is connected by pin joints, which are allowed
 to move along the \textit{z}-axis. Let the general coordinate of this $N$-DOF
 system be $\mathbf{q}={{\left[ \begin{matrix}
 {{\theta }_{1}} & \cdots  & {{\theta }_{j}} & \cdots  & {{\theta }_{N}}  \\
\end{matrix} \right]}^{T}}.
$
Given the identical initial angles imposed on the unit cells, the equation
of motion for this system can be expressed as
\begin{equation} \label{eq:origamimodel}
\mathbf{G}^T\mathbf{\hat{M}G\ddot{q}}+{{\mathbf{G}}^{T}}\mathbf{\hat{M}\dot{G}\dot{q}}={{\mathbf{G}}^{T}}{{\mathbf{f}}^{ex}}
\end{equation}
where
\begin{eqnarray*}
\mathbf{\hat{M}} &=& \textrm{diag}\left[ \begin{matrix}
   {{{\mathbf{\hat{M}}}}_{1}} & \cdots  & {{{\mathbf{\hat{M}}}}_{N}}  \\
\end{matrix} \right], \\
\mathbf{\hat{M}}_j &=& \textrm{diag}\left[ \begin{matrix}
   {m} & {m} & {J} & {m} & {m} & {J}  \\\end{matrix} \right],\ \ \ \ \ \ \ \ \ \  \\
{{\mathbf{G}}^{T}} &=& \left[ \begin{matrix}
   \mathbf{G}_{1}^{T} & {{\mathbf{O}}_{1\times 6}} & {{\mathbf{O}}_{1\times 6}} & \cdots  & \cdots  & \cdots  & {{\mathbf{O}}_{1\times 6}}  \\
   \mathbf{g}_{2}^{T} & \mathbf{G}_{2}^{T} & {{\mathbf{O}}_{1\times 6}} & \cdots  & \cdots  & \cdots  & {{\mathbf{O}}_{1\times 6}}  \\
   \vdots  & \vdots  & \vdots  & \vdots  & \vdots  & \vdots  & \vdots   \\
   \mathbf{g}_{j}^{T} & \mathbf{g}_{j}^{T} & \cdots  & \mathbf{G}_{j}^{T} & \mathbf{g}_{j}^{T} & \cdots  & {{\mathbf{O}}_{1\times 6}}  \\
   \vdots  & \vdots  & \vdots  & \vdots  & \vdots  & \vdots  & \vdots   \\
   \mathbf{g}_{N}^{T} & \mathbf{g}_{N}^{T} & \mathbf{g}_{N}^{T} & \cdots  & \cdots  & \cdots  & \mathbf{G}_{N}^{T}  \\
\end{matrix} \right], \\
\mathbf{G}_{j}^{T} &=& \left[ \begin{matrix}
   -3L\cos {{\theta }_{j}} & -L\sin {{\theta }_{j}} & 1 & -L\cos {{\theta }_{j}} & -L\sin {{\theta }_{j}} & -1  \\
\end{matrix} \right], \\
\mathbf{g}_{j}^{T} &=& \left[ \begin{matrix}
   -4L\cos {{\theta }_{j}} & 0 & 0 & -4L\cos {{\theta }_{j}} & 0 & 0 \\
\end{matrix} \right], \\
\mathbf{O}_{1\times 6} &=& \left[ \begin{matrix}
   0 & 0 & 0 & 0 & 0 & 0  \\
\end{matrix} \right].
\end{eqnarray*}
Also, $\mathbf{f}^{ex}$ is an external force vector defined as follows
\begin{equation}
	\mathbf{f}^{ex} = \left[ \begin{matrix}
   \left( \mathbf{f}_{1}^{ex} \right)^{T} & \cdots  & \left( \mathbf{f}_{j}^{ex} \right)^{T} & \cdots  & \left( \mathbf{f}_{N}^{ex} \right)^{T}  \\
\end{matrix} \right]^{T}
\end{equation}
where

\[\mathbf{f}_{j}^{ex} = \left\{ 
  \begin{array}{l l}
    [\begin{matrix} {F}^{ex}, & 0, & -2{{k}_{\theta }}\left( {{\theta }_{1}}-{{\theta }_{1,0}}\right)-{{F}^{ex}}L\cos {{\theta }_{1}}, \end{matrix} \\
     \quad \begin{matrix} 0, & 0, & -2{{k}_{\theta }}\left( {{\theta }_{1,0}}-{{\theta }_{1}} \right) \end{matrix}]^{T} & \, \text{if $j = 1$}\\
    
    [\begin{matrix} 0, & 0, & -2{{k}_{\theta }}\left( {{\theta }_{j}}-{{\theta }_{j,0}}\right), \end{matrix} \\
     \quad \begin{matrix} 0, & 0, & -2{{k}_{\theta }}\left( {{\theta }_{j,0}}-{{\theta }_{j}} \right) \end{matrix}]^{T} & \, \text{if $j = 2{\ldots}N$}\\  
  \end{array} \right.\]
  
%
%


\noindent See Supplemental Material for the details of this derivation~\cite{SI}. 

\subsection{Lumped Mass Model}
In this section, we introduce a lumped mass model, in which a chain of 
origami cells is modeled as lumped masses connected by nonlinear springs
(see Fig.~\ref{fig:Modeling_origami}(b)). The strain softening 
behavior of the TMP unit cell considered herein leads to the following power-law relationship:
\begin{equation} \label{eq:powerlaw}
F^{ex}=A{\delta}^n
\end{equation}
where $\delta$ is the compressive displacement, and the coefficient $A$ and the exponent $n$ are the constant values 
determined by curve fitting of Eq.~\eqref{eq:FS_curve}. 

Since the power-law relationship in Eq.~\eqref{eq:powerlaw} assumes only a positive displacement as an argument, we need to apply a displacement offset ($d_0$) towards the tension side, so that the lumped mass model can approximate the force-displacement curve of the multi-bar linkage model not only in the compressive region, but also in the tensile domain.
In Fig.~\ref{fig:SingleTMPcell}(c), the dashed curve shows the fitted power-law relationship for the multi-bar linkage model, 
where the black circle represents (along
the horizontal axis) the displacement offset $d_0$.

By using this simple force-displacement relationship, we can derive a general
expression of the equation of motion as follows:
\begin{equation} \label{eq:LumpedMassModel}
M\ddot{u}_j = A \left[ d_0 + \delta_{j-1,j} \right]_+^n - A\left[ d_0 + \delta_{j,j+1} \right]_+^n
\end{equation}
where $M$ is the lumped mass corresponding to $2m$, $n\in\mathbb{R}$, and the $[+]$ 
sign outside of the brackets indicates that we take only positive values of the strain $\delta_{j,j+1}=u_j-u_{j+1}$. Note that this form of equation has been 
used widely for analyzing nonlinear waves propagating in discrete systems
in the case of strain-hardening interactions (i.e., $n > 1$ in Eq.~\eqref{eq:LumpedMassModel}, 
e.g., granular crystals). Therein, the formation and propagation of nonlinear wave 
structures, such as solitary waves~\cite{Nesterenko,review_Sen} and discrete breathers~\cite{review_PK,review_Theo}, 
have been well studied. The interpretation of origami dynamics 
via this nonlinear lumped mass system opens up a broad, novel
potential vein of studies. Indeed, one advantage of modeling the origami lattice
in this way is that many tools and results obtained in the context of granular crystals
can be applied in our setting. 
For example, the recent work of~\cite{Herbold} examined
a one-dimensional discrete system under the power-law relationship
of strain-softening springs (i.e., $n < 1$ in Eq.~\eqref{eq:LumpedMassModel}). 
This study reported the propagation of rarefaction waves through dynamic simulations and 
a long wavelength approximation, where it was shown that the width of the rarefaction 
wave is independent of the wave speed.
Likewise, the analysis of nonlinear waves in post-buckled structures has been also attempted using a similar discrete system~\cite{Spadoni2014}.
%
In this article, we extend the theoretical results
 in such nonlinear-spring systems 
 by introducing a systematic tool for the computation
of numerically exact traveling waves, which will be discussed 
in Sec.~\ref{sec_IV}. We also address the subject
of their dynamical stability in the 
Supplemental Material~\cite{SI}.


\section{Numerical Simulations}\label{sec_III}
To examine the dynamic characteristics of the origami-based 
structure and compare the results from the two reduced models, we conduct
numerical computations of wave propagation under a compressive impact. Also, we apply various amplitudes 
of impact force to the multi-bar linkage model in order to examine the speed of
both compressive and tensile strain waves,
especially focusing on the dominant traveling wave.

\subsection{Waveform analysis} \label{sec:waveshape}
%

\begin{figure*}[htbp]
\centerline{ \epsfig{file=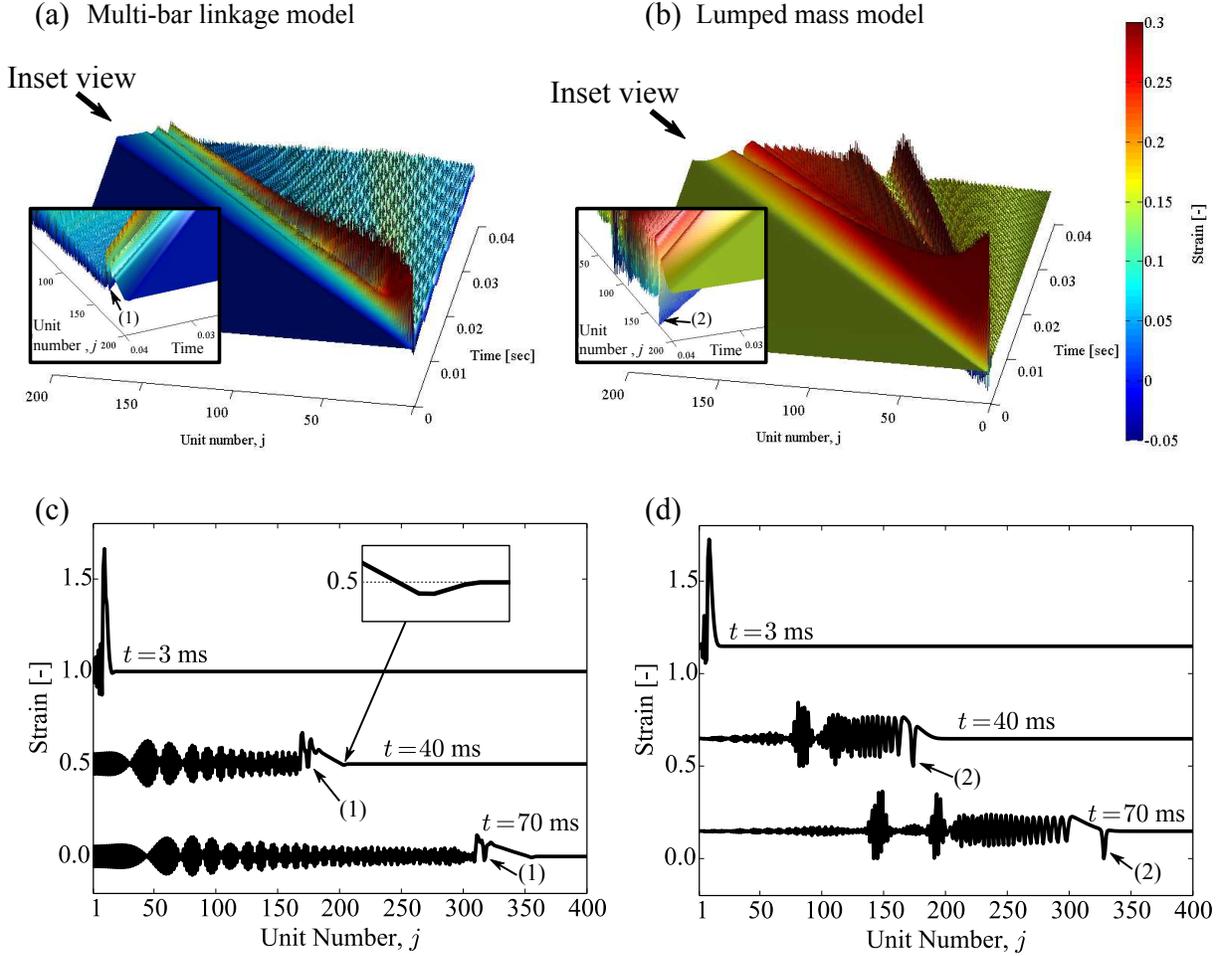,width=0.9 \textwidth} }
\caption{(Color online) Space-time contour plots of strain wave propagation based on 
\textbf{(a)} the Multi-bar linkage model and \textbf{(b)} 
the Lumped mass model. Insets show the magnified view of rarefaction waves.
Temporal plots of strain waves using \textbf{(c)} the Multi-bar linkage model and 
\textbf{(d)} the 
Lumped mass model. The inset in \textbf{(c)} shows the magnified view of the
leading edge. The arrows (1) and (2) point to the rarefaction wave
present in the dynamics.}
\label{fig:Sim_results_case2}
\end{figure*}
We perform numerical computations where a compressive impact is applied to
the first unit cell with the right end of the $N$-th unit cell kept fixed as shown
in Fig.~\ref{fig:Modeling_origami}(a). The strain waves propagating in a uniform
chain of $N=400$ unit cells are examined numerically. In the case of the multi-bar linkage model, the relative strain is defined as
\begin{equation} \label{eq:strain}
\eta_j = \frac{h_{j,0}-h_j}{h_{j,0}}
\end{equation}
where $h_j = 4L\sin{{\theta}_j}$ and $h_{j,0}=4L\sin{\theta_{j,0}}$ (see 
Fig.~\ref{fig:SingleTMPcell}(b)). The numerical constants used in the 
calculation are the following: $L=5\,\textrm{mm}$, $m=0.39\,\textrm{g}$, 
$k_{\theta}=1.0\,\textrm{Nm/rad}$, and $\theta_{j,0}=45^{\circ}$. To apply impact excitation, we impose $F^{ex}=100\,\textrm{N}$
for the first 1 $\textrm{ms}$ and $F^{ex}=0\,\textrm{N}$ after the first
1 $\textrm{ms}$ in our simulations. From the force-displacement curve based on
these constants, we obtain $n=0.64$ and $A=2,938\,\textrm{N/m}^n$, given an 
initial displacement offset of $d_0 = 2.1\,\textrm{mm}$ for the 
power-law 
approximation. 
In the case of the lumped mass model, the relative strain is defined as
\begin{equation} 
\eta_j = \frac{u_{j+1}-u_j}{d_0}.
\end{equation}

Figures~\ref{fig:Sim_results_case2}(a) and (b) show space-time contour plots
of strain wave propagation under compressive impact, while Figs.~\ref{fig:Sim_results_case2}(c)
and (d) show the strain waveforms corresponding to $t = 3, 40$, and $70 \,\textrm{ms}$. 
After the impact force is applied to the system, the first compressive impact
attenuates quickly as the strain waves propagate through the system, and then
a rarefaction wave appears in front of the first compressive wave (see the insets as well as the arrows
(1) and (2) in Fig.~\ref{fig:Sim_results_case2}). 
It should be also noted that due to the strain-softening behavior, the amplitude of the compressive force is reduced drastically as the wave propagates along the chain.
Since both the multi-bar linkage model and the
lumped mass 
model have this strain-softening nature, the same type of rarefaction waves is
observed.

In addition, the inset of Fig.~\ref{fig:Sim_results_case2}(c) shows the magnified 
view of the leading edge of the propagating strain wave. This leading wave is created due to the effect of inertia in the multi-bar linkage model.
That is, when the
first unit cell folds right after the compressive impact, the second unit cell is 
pulled by the first unit cell before the compressive force propagates 
to the next unit cell. Therefore, the tensile strain appears in front of the first compressive wave in the multi-bar linkage model.
Comparing the numerical results of the two models, the lumped mass model 
captures the multi-bar linkage model dynamics 
even quantitatively at short times,
while the agreement between the two becomes qualitative at longer 
time scales. 

Let us also note in passing that in the wake of this primary 
rarefaction pulse, 
we observe radiative dispersive wavepackets both in the  multi-bar linkage model
and in the lumped mass model. These wavepackets apparently travel
maximally with the speed of sound in the medium, 
while the 
rarefaction pulse outrunning them is apparently supersonic.
We will return to this point to corroborate it further by our
numerical bifurcation analysis in the next section.
Additionally, it should be noted that in the lumped mass model, highly
localized structures with a clear envelope can be discerned (see
e.g., the vicinity of unit number 150 of the $70$ ms panel of Fig.~\ref{fig:Sim_results_case2}(d)), which seem to have the form
of breather excitations, which are exponentially localized in space and periodic in time~\cite{flach, AubryDB_review}. A closer inspection of Fig.~\ref{fig:Sim_results_case2}(b)
also seems to suggest that such coherent wavepackets travel more
slowly than the dispersive radiation. The multi-bar linkage model also exhibits such time-periodic patterns, but there is no clear signature of spatial localization. While these nonlinear wave structures are worth investigating, this topic is beyond the scope of this paper, and we do not explore them further here.

\subsection{Wave speed analysis}
%
\begin{figure*}[htbp]
\centerline{ \includegraphics[width=.975\textwidth]{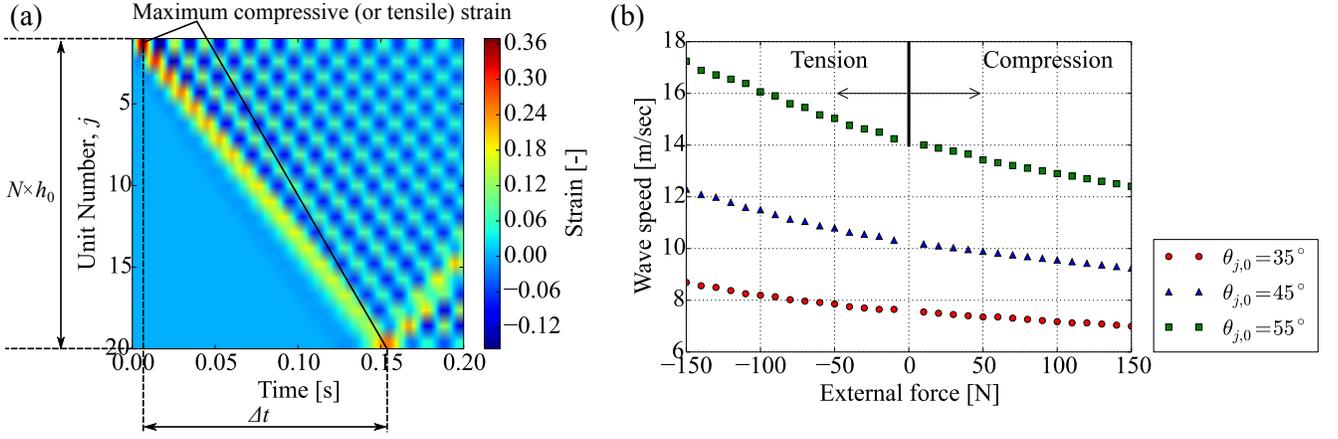}}
\caption{(Color online) \textbf{(a)} Surface map of strain field to calculate wave speed. \textbf{(b)} Wave speed of strain waves as a function of external force ranging from $-150\,\textrm{N}$ to $+150\,\textrm{N}$. Numerical simulations are
based on $L=25\,\textrm{mm}$, $m_j=19.7\,\textrm{g}$, $N = 20$ and $k_{\theta}=1.0\,\textrm{Nm/rad}$. }
\label{fig:wavespeed_calculation}
\end{figure*}
The propagation speed of strain waves is now investigated numerically under various 
amplitudes of impact force. The wave speed is approximated as follows
\begin{equation} \label{eq:wavespeed}
V_{\epsilon}=\frac{Nh_0}{{\Delta}t}
\end{equation}
where $h_0$ is the initial height of the unit cell, and ${\Delta}t$ is the time
span in which the strain wave propagates from the first unit cell to the $N$-th 
unit cell (see Fig.~\ref{fig:wavespeed_calculation}(a)). 
The propagating wave speeds calculated are depicted in Fig.~\ref{fig:wavespeed_calculation}(b) under three 
different initial folding angles: $\theta_{j,0} = 35^{\circ}, 45^{\circ}$, and
$55^{\circ}$. 
It is evident that the wave speed is 
altered by the impact 
force, which is one of characteristics of nonlinear waves. However, it should be noted
that in the compressive regime, the wave speed decreases as the compressive impact
increases. This is in sharp contrast to conventional nonlinear waves formed in the system
of strain-hardening lattices~\cite{Nesterenko,review_Sen}.
A different trend is observed in the tensile regime, where the wave
speed increases as the tensile impact increases. It is also noteworthy that the wave
speed curve can be shifted by changing the initial folding angle. Therefore, we can
control the speed of the waves propagating through the origami-based metamaterials
by altering their geometrical configurations, implying their inherent dynamical tunability.

\section{Exact rarefaction waves of the lumped mass model}\label{sec_IV}

We now turn our attention to a more systematic analysis and
understanding of the rarefaction waves in the simpler lumped mass
model; notably, our conclusions here in that regard are of broader
interest to previously discussed settings such as those 
of~\cite{Herbold,fratern}.
Based on the previous analysis, we numerically investigate the existence
and dynamical stability of exact rarefaction waves of the lumped mass model 
[cf. Eq.~(\ref{eq:LumpedMassModel})]. In particular, we consider the model
in the strain variable $\delta_{j,j+1}$ written 
as
\begin{eqnarray} 
M \ddot{\delta}_{j,j+1}&=&A\lbrace\left[d_{0} + \delta_{j-1,j}\right]_{+}^{n}%
-2[d_{0}+\delta_{j,j+1}]_{+}^{n}\nonumber \\
&+&\left[d_{0}+ \delta_{j+1,j+2}\right]_{+}^{n}\rbrace.
\label{strain_model}
\end{eqnarray}
The existence 
and the spectral 
stability of traveling waves of Eq.~(\ref{strain_model}) with wave 
speed $c$ must be examined through the ansatz $\delta_{j,j+1}(t)=\delta(j-c\,t):=\Phi(\xi,t)$, i.e., going to the co-traveling wave frame where the
relevant solution appears to be steady and hence amenable to a 
spectral stability analysis.
Then, $\Phi$ solves the advance-delay differential equation
\begin{widetext}
\begin{equation}
 {\Phi}_{tt}(\xi,t)=-c^{2} {\Phi}_{\xi\xi}(\xi,t)+2c%
 {\Phi}_{\xi t}(\xi,t)+\frac{A}{M}\Big\lbrace\left[d_{0}+ {\Phi}(\xi-1,t)\right]_{+}^{n}-%
2\left[d_{0}+ {\Phi}(\xi,t)\right]_{+}^{n}+\left[d_{0}+ {\Phi}(\xi+1,t)\right]_{+}^{n}\Big\rbrace.
\label{step_2}
\end{equation}
\end{widetext}

Traveling waves of Eq.~\eqref{strain_model} correspond to stationary (time 
independent) solutions  $\Phi(\xi,t) = \phi(\xi)$ of Eq.~\eqref{step_2},
satisfying
\begin{eqnarray} 
0 &=& - c^{2}\phi_{\xi \xi} + \frac{A}{M}\lbrace \left[d_{0}+\phi(\xi-1)\right]_{+}^{n}%
-2\left[d_{0}+\phi(\xi)\right]_{+}^{n}\nonumber\\
            &+&\left[d_{0}+\phi(\xi+1)\right]_{+}^{n}\rbrace.
\label{adv-del}
\end{eqnarray}
To obtain numerical solutions of Eq.~\eqref{adv-del}, we employ a 
uniform spatial discretization of $\xi$ consisting of $l$ points 
$\xi_{k}$ ($k=-\frac{l-1}{2},\dots,0,\dots,\frac{l-1}{2}$) with lattice spacing $\Delta{\xi}$ chosen
such that $q=1/\Delta{\xi}$ is an integer. Then, the field $\phi(\xi)$
is replaced by its discrete counterpart, i.e., $\phi_{k}:=\phi(\xi_{k}) = \phi(k \Delta{\xi})$. 
The second-order spatial derivative appearing in Eq.~(\ref{adv-del}) 
is replaced by a modified central difference approximation
$(\phi_{k-2}-2\phi_{k}+\phi_{k+2})/(4\Delta{\xi}^2)$. The reason for 
this choice of central difference  is connected to the stability 
calculation to be discussed in the Supplemental Material~\cite{SI}. 
Using this discretization, 
Eq.~(\ref{adv-del}) becomes the following root-finding problem,
\begin{eqnarray}
0&=&-c^{2}\frac{\phi_{k-2} - 2\phi_{k}+\phi_{k+2}}{ 4\Delta{\xi}^2}%
+\frac{A}{M}\lbrace \left[d_{0}+\phi_{k-q}\right]_{+}^{n} \nonumber \\
&-& 2\left[d_{0}+\phi_{k}\right]_{+}^{n}+\left[d_{0}+\phi_{k+q}\right]_{+}^{n}\rbrace
\label{disc-adv-del}
\end{eqnarray}
which is solved via Newton iterations. We employ periodic boundary 
conditions at the edges of the spatial grid. 
We are interested specifically in rarefaction
waves, and thus we use the profiles obtained via the numerical simulations
of Sec.~\ref{sec:waveshape} to initialize the Newton solver, see e.g. 
arrow (2) of Fig.~\ref{fig:Sim_results_case2}(d). Herein, we consider an origami lattice with 
$L = 25\,\textrm{mm}$, $k_\theta = 1.0\,\textrm{Nm/rad}$ and $\theta = 55^\circ$. 
The corresponding best-fit values of the parameters of the lumped-mass model
are $A=280\,\textrm{N}/\textrm{m}^n$, $n=0.53$, $m=19.7\,\textrm{g}$ with 
$M=2m$ and $d_{0}=12\,\textrm{mm}$. 

\begin{figure*}[htbp]
\begin{center}
\vspace{-0.1cm}
\mbox{\hspace{-1cm}
\subfigure[][]{\hspace{-0.2cm}
\includegraphics[height=.19\textheight, angle =0]{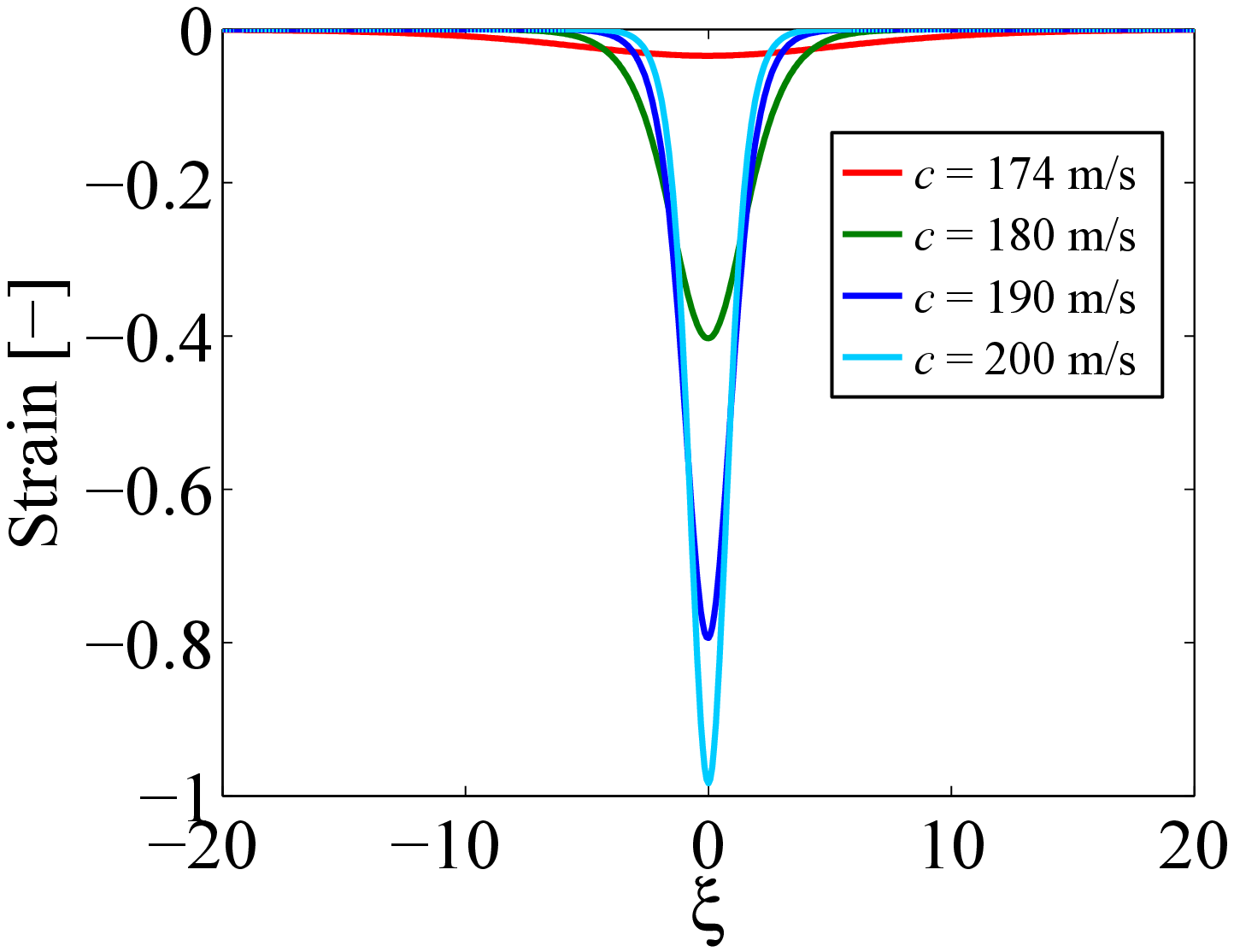}
\label{fig6a}
}
\subfigure[][]{\hspace{-0.2cm}
\includegraphics[height=.19\textheight, angle =0]{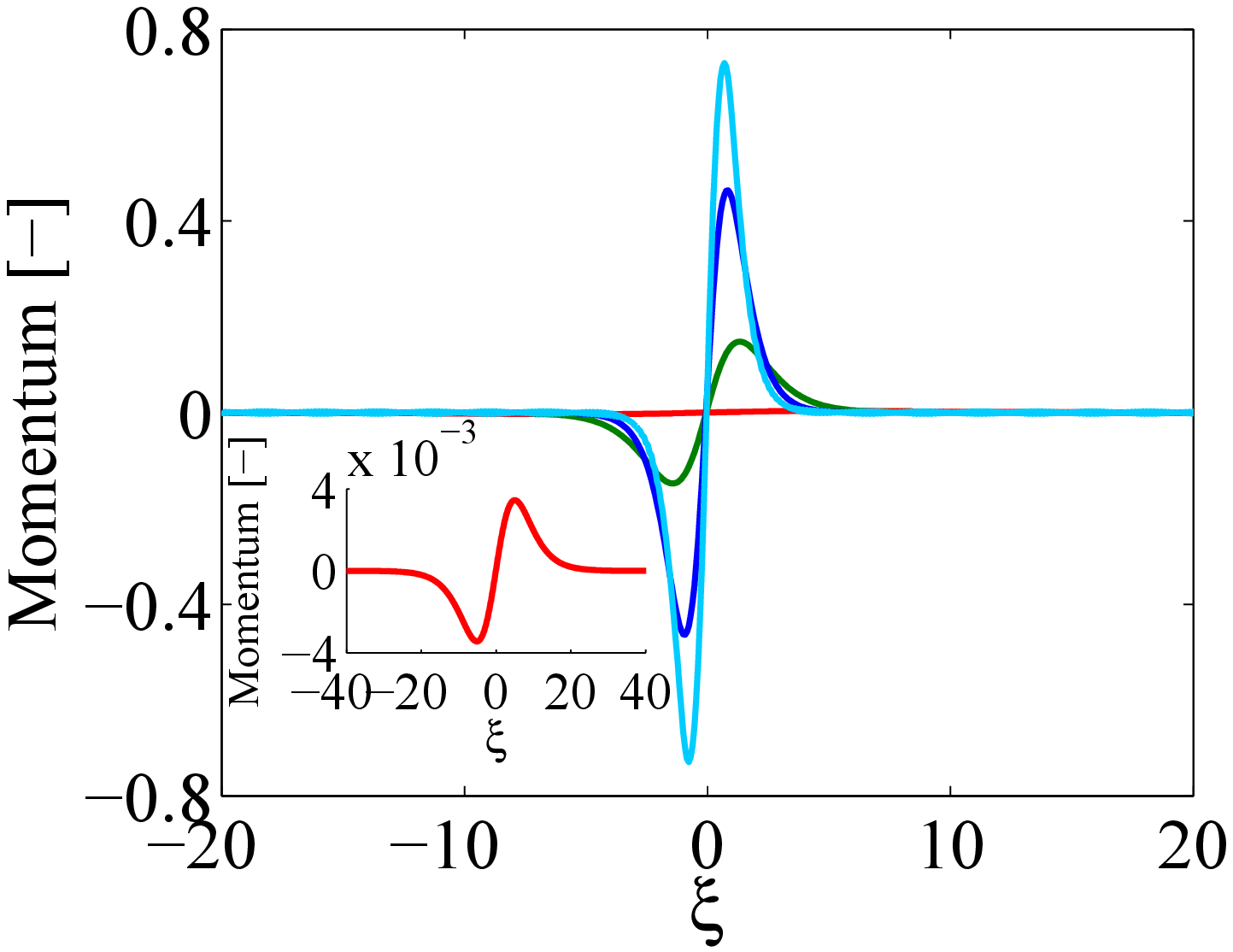}
\label{fig6b}
}
\subfigure[][]{\hspace{-0.2cm}
\includegraphics[height=.19\textheight, angle =0]{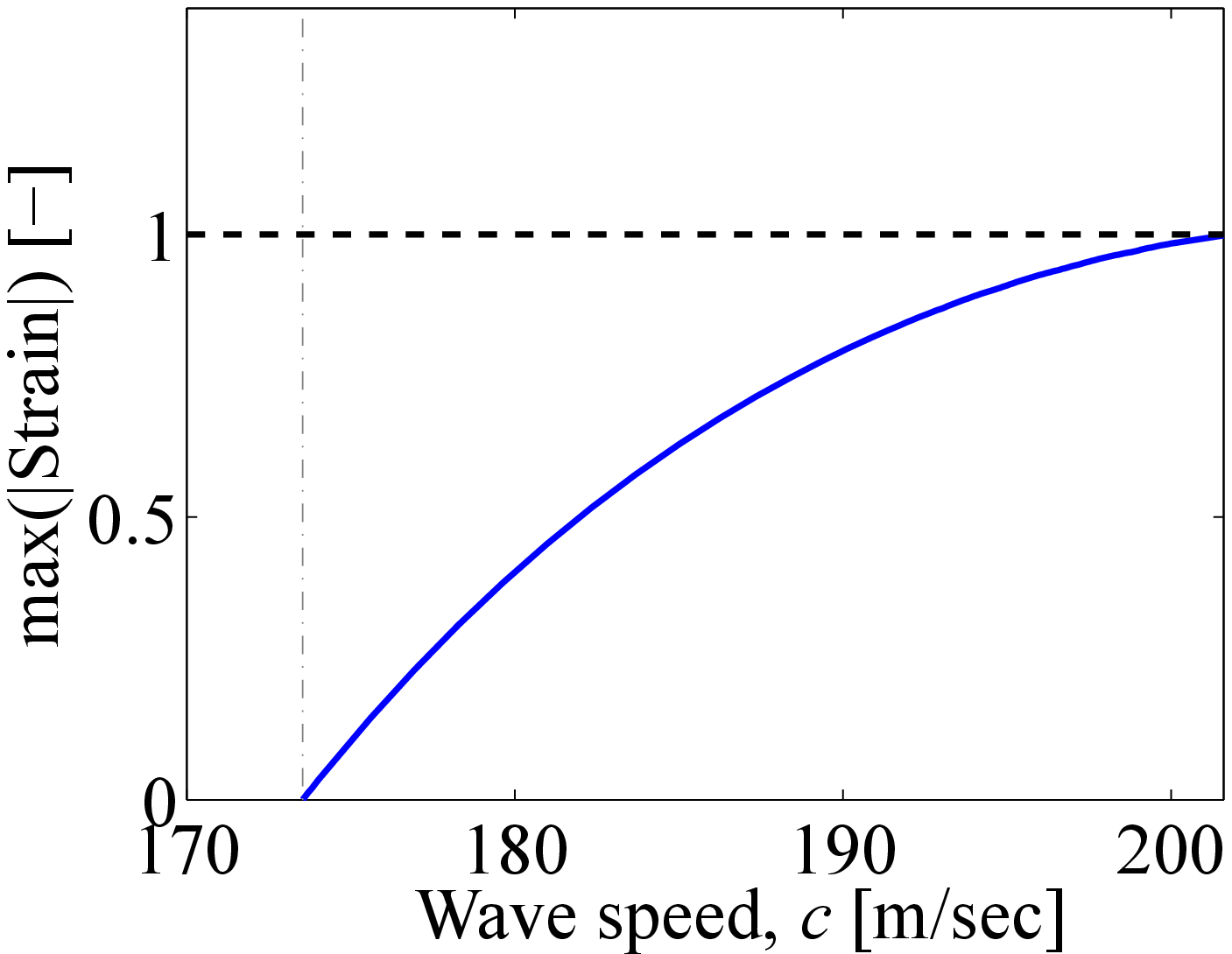}
\label{fig6c}
}
}
\end{center}
\caption{(Color online) Summary of numerical results on continuations of
rarefaction waves over wave speed $c$ with $l=4001$ points and $\Delta{\xi}=1/13$:
\textbf{(a)} Relative strain profiles for various values of the wave speed $c$.
\textbf{(b)} Relative momenta corresponding to (a). \textbf{(c)}
Maximum of the absolute value of the relative strain variable as
a function of the wave speed. Note that the horizontal dashed black line 
corresponds to the value of pre-compression in normalized units (or, equivalently,
$d_{0}$ in physical units), while the vertical dashed-dot gray line 
corresponds to the value of the speed of sound $c_s$ of the medium.}
\setlength{\abovecaptionskip}{-20pt}
\label{fig:continuations_pbc}
\end{figure*}

In Fig.~\ref{fig:continuations_pbc}, numerically exact 
rarefaction waves (i.e., solutions of Eq.~\eqref{disc-adv-del} 
with a prescribed
tolerance) are presented for various values of the wave speed $c$. In 
particular, Fig.~\ref{fig6a} shows the rarefaction waves in terms of the
relative strain variable  $\phi/d_0$, while Fig.~\ref{fig6b} shows the corresponding
relative momenta  $\phi'/d_0$. Note that the tails decay to zero monotonically, 
implying that the traveling structure does not resonate with the linear modes of
the system, as the wave is supersonic. It is not surprising then that our 
parametric continuation in the 
wave speed $c$ reveals a critical minimum value 
$c_s=\sqrt{nAd_{0}^{n-1}/M} = 173.5$ m/s, 
which is the sound speed of the chain (see the vertical dashed-dot gray line of Fig.~\ref{fig6c}).
This is consistent with the long-wavelength analysis of \cite{Herbold}
and also with our observations of the previous section indicating
that the wave outruns the small amplitude radiation tails behind it. 
Thus, similarly to systems with $n>1$ \cite{Nesterenko,Stefanov}, the 
rarefaction waves
of the origami lattice are traveling faster than any linear waves of the system.
However, in contrast to solitary waves in systems with $n>1$, the amplitude of the
rarefaction waves in the origami system have a natural bound determined by the
precompression $d_0$ of the system, in which case the particles come out of 
contact
(see the horizontal dashed black line of Fig.~\ref{fig6c}). Although 
waves with
amplitude exceeding this value are in principle possible, we were unable to 
identify any ones such
numerically. An interesting open problem would be to prove rigorously if such a bound
exists. Another interesting related 
problem is if there is a critical maximum value of $c$. 
Our numerical continuation algorithm did indeed terminate due to 
lack of convergence at
$c\approx201.6 $ m/s, but this could have been a result of the ill-conditioned 
nature of the 
Jacobian matrix as the amplitude approached the critical limit of $d_0$.

The robustness of a solution $\phi^0$  of Eq.~(12) can be investigated through a spectral stability analysis.
To that end, we substitute the linearization ansatz $\Phi(\xi,t) = \phi^0 + \epsilon a(\xi) e^{\lambda t}$
into Eq.~(11), which yields an eigenvalue problem at order $\epsilon$ (see 
the Supplemental Material~\cite{SI}).
We considered solutions at various wave speeds $c$ and found in each case at least one eigenvalue with a small real part, indicating
a (very weak) instability. However, the eigenvalues are highly sensitive to e.g. lattice size and choice of discretization, suggesting
that these instabilities may 
be ``spurious".  To check this, we performed dynamical
simulations of the perturbed rarefaction waves at the level of Eq.~\eqref{strain_model} and found that they are all
robust against small perturbations (see Fig. 8 of the Supplemental Material~\cite{SI}). This suggests that the very weak
instabilities predicted via the spectral stability analysis are indeed spurious.
While a heuristic argument for the presence of spurious instabilities is provided in the Supplemental Material~\cite{SI},
the construction of a mathematically consistent algorithm for the computation of eigenvalues in this context
remains an important open problem.

\section{Conclusions \& Future Challenges}\label{sec_V}

 In the present work, we investigated nonlinear wave dynamics in 
origami-based metamaterials
 consisting of building blocks based on 
Tachi-Miura polyhedron (TMP) cells. We analyzed the kinematics
 of the TMP unit cell using a simple multi-bar linkage model and found that
 it exhibits tunable strain-softening behavior under compression due to its
 geometric nonlinearity. We observed that upon impact, this origami-based 
 structure supports the formation and propagation of rarefaction waves.
The resulting evolution
 features 
a tensile wavefront despite the application of compressive impact. 
A further reduction was also offered based on the fitted force-displacement
formula for a single cell, in the form of a lumped mass model.
 In the latter case we obtained numerically exact rarefaction
 waves and studied their spectral and especially dynamical stability.
 The dynamical features observed herein may constitute  
a highly useful feature towards the efficient 
mitigation of impact by converting
 compressive waves into rarefaction waves and disintegrating high-amplitude 
 impulses into small-amplitude oscillatory wave patterns. We also demonstrated 
the potential tunability
 of the wave speed by altering initial folding conditions of the origami-based 
 structure, which naturally opens up the feasibility of controlling stress wave
 propagation in an efficient manner. 

The rather unique nonlinear wave dynamics of origami 
 structures can lead to a wide range of applications, such as 
tunable wave transmission
 channels and deployable impact mitigating layers for space and other engineering
 applications. On the theoretical/computational side, there is also a large number of intriguing questions that
are emerging. For one, a more detailed comparison of the coherent
structure propagation in the multi-bar linkage model vs. that of
the lumped-mass model would be an interesting topic for further consideration. 
This would help uncover the dynamical features
leading to the apparent weak amplitude decay in the former, while the
latter contains robust solutions and sustained long-time propagation. %
Still at the single wave level, an exploration of the delicate issues
of spectral stability by means of different numerical methods and
of the corresponding dynamical implications would be of particular interest.
Subsequently, understanding further the dynamics and interactions of
multiple rarefaction wave patterns would also be a relevant theme
for future investigations. These topics are currently under 
active consideration and will be reported in future publications.

\begin{acknowledgments}
J.Y. acknowledges the support of NSF (CMMI-1414748) and ONR (N000141410388).
E.G.C. and P.G.K acknowledge support from the US-AFOSR under grant FA9550-12-10332.
P.G.K. also acknowledges support from the NSF under grant 
DMS-1312856, from ERC and FP7-People under grant 605096, and from the Binational (US-
Israel) Science Foundation through grant 2010239. P.G.K.'s work at Los Alamos is 
supported in part by the U.S. Department of Energy. The work of C.C. was partially supported
by the ETH Zurich Foundation through the Seed Project ESC-A 06-14.
\end{acknowledgments}


\end{document}